\documentclass[reprint,aps,prl,twocolumn,superscriptaddress,showpacs,preprintnumbers,amsmath,amssymb,floatfix,merge,nofootinbib]{revtex4-1}
\usepackage{graphicx}  
\usepackage[mathlines]{lineno}
\usepackage{dcolumn}   
\usepackage{bm}        
\usepackage{amssymb}   
\usepackage{xcolor}
\usepackage{verbatim}
\usepackage[colorlinks,linkcolor=blue,citecolor=blue,urlcolor=blue]{hyper ref}

\begin{document}


\title{Constraints on Lightly Ionizing Particles from CDMSlite}

\author{I.~Alkhatib} \affiliation{Department of Physics, University of Toronto, Toronto, ON M5S 1A7, Canada}
\author{D.W.P.~Amaral} \affiliation{Department of Physics, Durham University, Durham DH1 3LE, UK}
\author{T.~Aralis} \affiliation{Division of Physics, Mathematics, \& Astronomy, California Institute of Technology, Pasadena, CA 91125, USA}
\author{T.~Aramaki} \affiliation{SLAC National Accelerator Laboratory/Kavli Institute for Particle Astrophysics and Cosmology, Menlo Park, CA 94025, USA}
\author{I.J.~Arnquist} \affiliation{Pacific Northwest National Laboratory, Richland, WA 99352, USA}
\author{I.~Ataee~Langroudy} \affiliation{Department of Physics and Astronomy, and the Mitchell Institute for Fundamental Physics and Astronomy, Texas A\&M University, College Station, TX 77843, USA}
\author{E.~Azadbakht} \affiliation{Department of Physics and Astronomy, and the Mitchell Institute for Fundamental Physics and Astronomy, Texas A\&M University, College Station, TX 77843, USA}
\author{S.~Banik}\email[]{samir.banik@niser.ac.in} \affiliation{School of Physical Sciences, National Institute of Science Education and Research, HBNI, Jatni - 752050, India}
\author{D.~Barker} \affiliation{School of Physics \& Astronomy, University of Minnesota, Minneapolis, MN 55455, USA}
\author{C.~Bathurst} \affiliation{Department of Physics, University of Florida, Gainesville, FL 32611, USA}
% ...




\collaboration{SuperCDMS Collaboration}
\noaffiliation

\date{\today}

\begin{abstract}
The Cryogenic Dark Matter Search low ionization threshold experiment (CDMSlite) achieved efficient detection of very small recoil energies in its germanium target, resulting in sensitivity to Lightly Ionizing Particles (LIPs) in a previously unexplored region of charge, mass, and velocity parameter space. We report first direct-detection limits calculated using the optimum interval method on the vertical intensity of cosmogenically-produced LIPs with an electric charge smaller than $e/(3\times10^5$), as well as the strongest limits for charge $\leq e/160$, with a minimum vertical intensity of $1.36\times10^{-7}$\,cm$^{-2}$s$^{-1}$sr$^{-1}$ at charge $e/160$. These results apply over a wide range of LIP masses (5\,MeV/$c^2$ to 100\,TeV/$c^2$) and cover a wide range of $\beta\gamma$ values (0.1 -- $10^6$), thus excluding non-relativistic LIPs with $\beta\gamma$ as small as 0.1 for the first time. \end{abstract}
\maketitle

\textit{Introduction}: The strong CP problem~\cite{cheng1988strong}, observation of neutrino oscillations~\cite{aguilar2001evidence}, matter-antimatter asymmetry~\cite{dine2003origin}, evidence for dark matter~\cite{darkmatterevidence}, and evidence for dark energy~\cite{sahni2014model} all suggest that the Standard Model (SM) provides an incomplete framework and motivate searches for physics beyond the SM. A promising avenue of exploration is the search for particles with a fractional electric charge. Fractionally-charged particles (FCPs) have charge $q$\,=\,$\pm$\,$fe$, where $e$ is the elementary charge and $f$ has a value between 0 and 1. Many extensions to the SM~\cite{abel2008kinetic,schellekens1990electric, composite_LIPs,chun1995models,glashow1961gauge} contain unconfined (`free') FCPs. A non-relativistic FCP has been proposed to explain the annual modulation signal observed by the DAMA/LIBRA~\cite{DAMA} and CoGeNT~\cite{CoGeNT} detectors~\cite{annualMod, annualMod2}. If particles with fractional charge exist, the lightest FCP must be stable, motivating these searches~\cite{shiu2013millicharged}.

\begin{figure}[!ht]
	\begin{center}
		\includegraphics[width=0.98\linewidth]{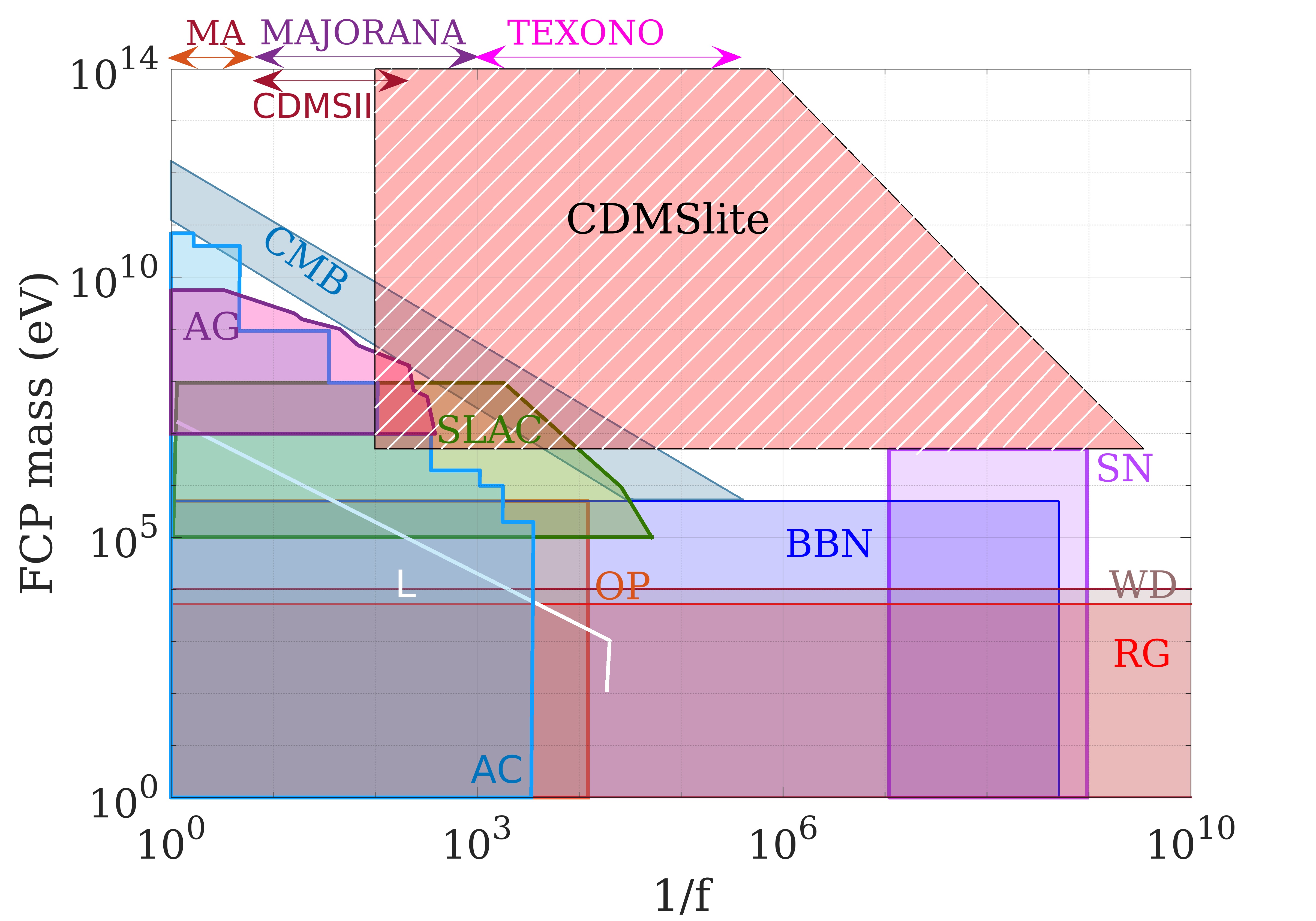}
		\caption{Constraints on FCP mass-charge parameter space from astrophysical observations and direct laboratory experiments. Direct-detection experiments MACRO (MA)~\cite{MACRO}, CDMS II~\cite{cdmsii-lips}, \textsc{Majorana}~\cite{Majorana}, TEXONO~\cite{texono} and CDMS\-lite (this search) constrain the intensity of cosmogenic FCPs; other constraints are adapted from Refs.~\cite{BBN,argoneutharnik2019millicharged} and include those from accelerator-based experiments (AC)~\cite{abbiendi2003search,beam_dump},
		ArgoNeut (AG)~\cite{argoneutharnik2019millicharged}, the search for the invisible decay of ortho-positronium (OP)~\cite{opos}, the SLAC millicharged particle search (SLAC)~\cite{prinz1998search}, the Lamb shift (L)~\cite{lamb-shift}, Big Bang nucleosynthesis (BBN)~\cite{BBN}, plasmon decay in red giants (RG)~\cite{white_dwarf_red}, plasmon decay in white dwarfs (WD)~\cite{white_dwarf_red}, the cosmic microwave background (CMB)~\cite{CMB_milli} and Supernova 1987A (SN)~\cite{supernovae}. The CDMSlite experimental constraints extend to the greatest value of $f^{-1}$ permitted such that the cosmological density of relic FCP does not exceed the total density of our universe~\cite{book:Kolb}. The constraints shown are for $\beta\gamma$ (see definition in text) of 0.1 which gives the least restrictive upper bounds on masses. This analysis is the first direct detection experiment to probe the impact of mass on the signal model. 
		}
		\label{parameter}
	\end{center}
\end{figure}

Constraints on FCP parameter space arise from astrophysical observations and laboratory experimentation~\cite{perl2009search,FCP_neutrino}. Figure~\ref{parameter} shows the constraints for free FCPs in the mass-charge plane. Free FCPs with small electric charge are known as Lightly Ionizing Particles (LIPs), because their mean energy loss per unit length $\langle\mathrm{d}E/\mathrm{d}x\rangle$ is suppressed as $f^2$~\cite{PDG} compared to particles with electron charge. Direct-detection experiments for LIPs are of particular interest, because they are sensitive to cosmogenically produced LIPs with both smaller $f$ and larger mass than any other experimental searches (see Fig.~\ref{parameter}).

Based on data from the SuperCDMS experiment, this paper describes the first direct search for LIPs with a variety of
incident $\beta\gamma$ values (0.1--10$^{6}$) and for $f$ as small as 10$^{-8}$, where $\beta$\,=\,$v/c$, $\gamma$\,=\,1$/\sqrt{1-\beta^2}$, and $v$ is the LIP velocity.
This is the first work to set limits on non-relativistic LIPs with $\beta\gamma$ as small as 0.1 (still $\sim$55 times larger than that expected of galactically bound~\cite{escape_velocity} LIPs). The analysis described herein searches for LIPs in an unexplored parameter space for masses between 5\,MeV$/c^2$ and 100\,TeV$/c^2$.

\textit{Experimental setup and data}: The SuperCDMS experiment employed five vertical stacks of detectors in the Soudan Underground Laboratory with each stack comprised of three germanium detectors~\cite{CDMSlite}. Each detector was a $\sim$600\,g cylindrical crystal with a 3.8\,cm radius and 2.5\,cm height, instrumented on each face (top and bottom) with four phonon and two ionization sensors. One of the detectors located in the middle of a stack was operated in CDMS\-lite mode~\cite{CDMSlite}, with a bias of 70\,V applied between its two faces. All others were biased at 4\,V. The detector operated at higher bias voltage amplifies the phonon signal via the Neganov-Trofimov-Luke (NTL) effect~\cite{neganov1985ussr}, allowing it to achieve a $<$\,100\,eV energy threshold. For information about detector operation, readout, and response, see Ref.~\cite{CDMSlite}.

This analysis uses data from the first period (February through July 2014) of the second CDMS\-lite run~\cite{CDMSlite}. CDMS\-lite Run 2 Period 1 had a live time of 97.81\,days, which was 84.6\,\% of the full Run 2 live time. Using only Period 1 data simplified the analysis with only a marginal reduction in sensitivity. This analysis was performed in an effectively ``blind'' fashion: although the CDMS\-lite Run 2 spectrum based on both Period 1 and Period 2 data is published~\cite{CDMSlite}, the analyzers did not use the Period 1 data to develop the limit-setting framework, including selection criteria, or to project sensitivities. Reconstructed energy depositions in the CDMS\-lite detector between 100\,eV and 2\,keV were analyzed with the 2\,keV upper limit chosen for the same reason as the CDMSlite Run 2 WIMP search~\cite{CDMSlite}. Energy-deposition spectra were simulated using the CDMS\-lite Run 2 background model~\cite{titrium} and were used to develop the analysis framework and make limit projections.

\textit{Signal model}: The LIP flux is attenuated by the atmosphere and rock overburden before reaching the experimental site. This can introduce an angular dependence in the LIP distribution. As in Ref. \cite{Majorana}, we consider two limiting cases: 1) an isotropic angular distribution and 2) a $\cos^2\theta$ angular distribution, where $\theta$ is the angle of an incident LIP relative to zenith. The former case corresponds to minimal attenuation for small $f$, while the latter case corresponds to muon-like attenuation for large $f$.

Expected energy-deposition probability distribution functions for LIPs passing through the CDMS\-lite detector are obtained using \textsc{Geant4}~\cite{AGOSTINELLI2003250} simulations. The simulation incorporates several processes including ionization, bremsstrahlung, pair production, and scattering (single and multiple). The \textsc{Geant4} Photo Absorption Ionization (\texttt{G4PAI}\footnote{We also use the \texttt{G4MuPairProduction}, \texttt{G4MuBremsstrahlung}, \texttt{G4WentzelVI}, and \texttt{G4eCoulombScattering} models with \texttt{G4MuPairProduction} and \texttt{G4MuBremsstrahlung} modified to respect the (fractional) electric charge of incident LIPs.}) \cite{G4PAI} model is used for the simulation of energy loss via ionization. The \texttt{G4PAI} model is typically used to model energy depositions in situations where a paucity of interactions is expected. 

\begin{figure}
	\begin{center}
                \includegraphics[width=0.95\linewidth]{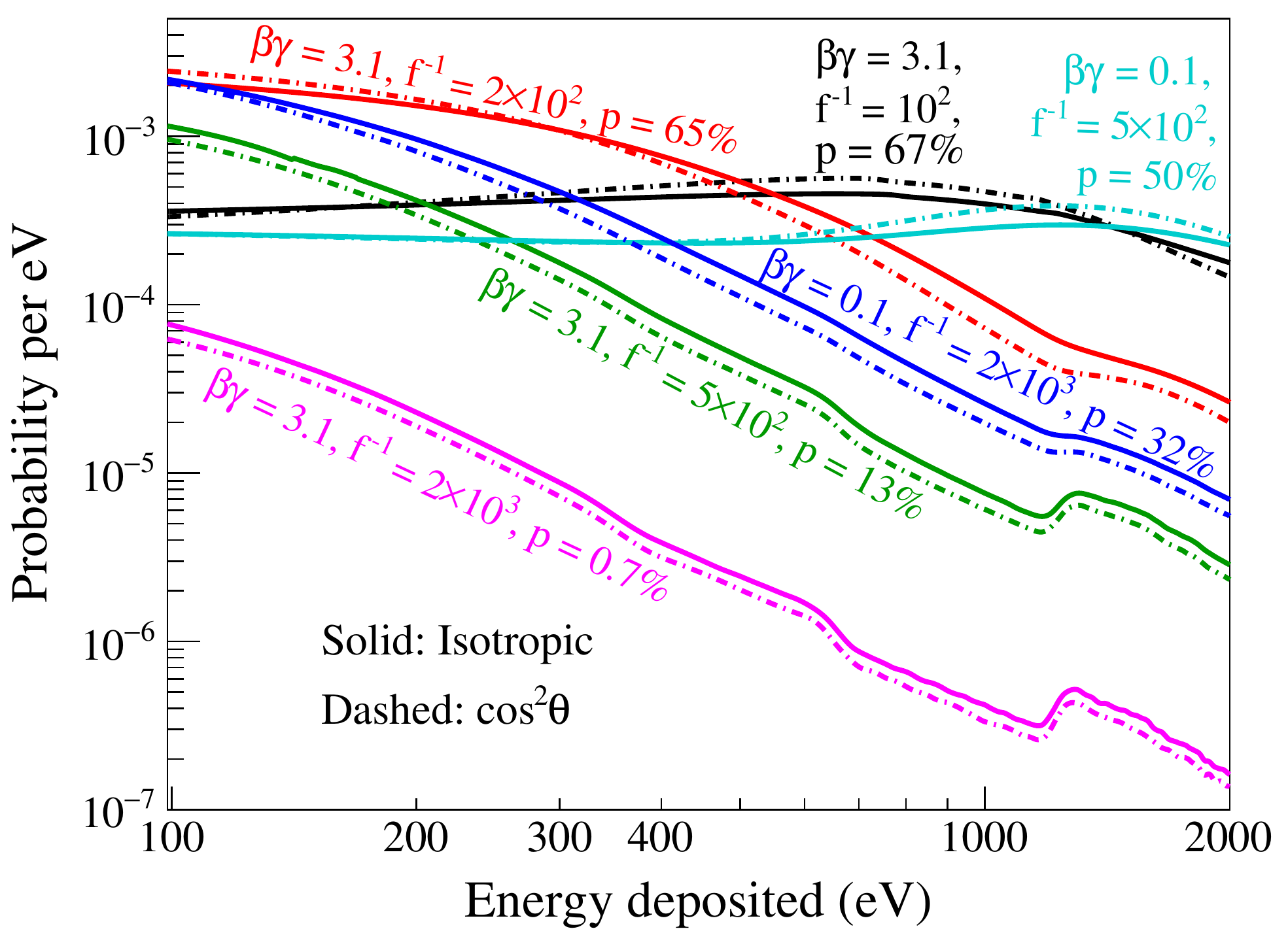}
		\caption{Simulated energy-deposition distributions averaged over incident angle $\langle\frac{\mathrm{d}P}{\mathrm{d}E}(f,\beta\gamma)\rangle$ for LIPs incident on the detector with two different values of $\beta\gamma$ and various $f^{-1}$ between 10$^{2}$ and $2\times10^{3}$, before the application of selection criteria, and after convolution with the detector energy resolution. The solid lines show the energy-deposition distributions assuming an isotropic incident LIP distribution, and the dotted lines show the distribution assuming a cos$^2\theta$ incident distribution. The figure also shows the total probability ($p$) of energy deposition  within the analysis energy range for the isotropic distribution. The atomic L-shell peaks at 1.3\,keV can be seen. For $f^{-1}>2\times10^{3}$, the shape of $\langle\mathrm{d}P/\mathrm{d}E\rangle$ does not change but merely scales down as $f^2$. The distributions are independent of mass in the range considered.} 
		\label{pdf}
	\end{center}
\end{figure}

The simulated energy-deposition distributions are convolved with the detector resolution~\cite{CDMSlite} and are calculated for a range of values of the LIP parameters: $f$, mass, and $\beta\gamma$ for both angular distributions. Figure~\ref{pdf} shows convolved energy-deposition distributions $\langle\mathrm{d}P/\mathrm{d}E\rangle$ for various $f$ and $\beta\gamma$ of LIPs incident on the detector. Example distributions for both minimum-ionizing ($\beta\gamma$\,$\sim$\,3.1) and non-relativistic ($\beta\gamma$\,$\sim$\,0.1) LIPs are shown to illustrate the impact of LIP velocity on the scattering probability for a given fractional charge.

 The LIP mass impacts the expected energy-deposition distribution through the bremsstrahlung process. The number of bremsstrahlung interactions is proportional to the inverse square of the LIP mass. Simulations show that the bremsstrahlung contribution to $\langle\mathrm{d}P/\mathrm{d}E\rangle$ is negligible within the analyzed energy window and the chosen LIP mass range\footnote{The LIP signal distribution is modeled for heavy charged particles with mass much greater than the mass of the electron. We did not calculate energy-deposition distributions for masses above 100\,TeV$/c^2$.} (5\,MeV$/c^2$ -- 100\,TeV$/c^2$); consequently, $\langle\mathrm{d}P/\mathrm{d}E\rangle$ is found to be effectively independent of the LIP mass in our analysis.

However, $\langle\mathrm{d}P/\mathrm{d}E\rangle$ is dependent on $\beta\gamma$ due to the ionization process, which is the dominant LIP energy-loss mechanism in the detector.  The ionization cross section is a function of $f$ and $\beta\gamma$. The assumption of minimum-ionizing ($\beta\gamma$\,$\sim$\,3.1) LIPs leads to the least restrictive limits for LIPs with $f^{-1}\gtrsim550$ as will be shown later. LIPs with smaller $f^{-1}$ and/or smaller $\beta\gamma$ ($\lesssim$ 1) have substantial probability of depositing energy above the largest energy deposition considered (2\,keV), resulting in a reduced LIP sensitivity. This dependence of LIP sensitivity on $\beta\gamma$ motivates our consideration of a range of LIP $\beta\gamma$ values.

\textit{Selection criteria and efficiency}: All data selection criteria used in the CDMS\-lite Run 2 WIMP search~\cite{CDMSlite} including the single-detector-hit criterion and the fiducial-volume criterion are applied to this LIP search. 
However, for LIPs the efficiencies of the single-detector-hit and fiducial-volume selections tend to be lower than those for WIMPs; correction factors to account for these relative inefficiencies are calculated using Monte Carlo simulation.
The product of efficiency correction factors for the single-detector-hit  ($\epsilon_{\tiny{\mbox{sdh}}}(f,\beta\gamma)$) and fiducial-volume ($\epsilon_{\tiny{\mbox{fv}}}(f,\beta\gamma)$) criteria is taken as the combined efficiency correction factor ($\epsilon_{\tiny{\mbox{corr}}}(f,\beta\gamma)$).

\begin{figure}[t!!]
	\begin{center}
		\includegraphics[width=0.95\linewidth]{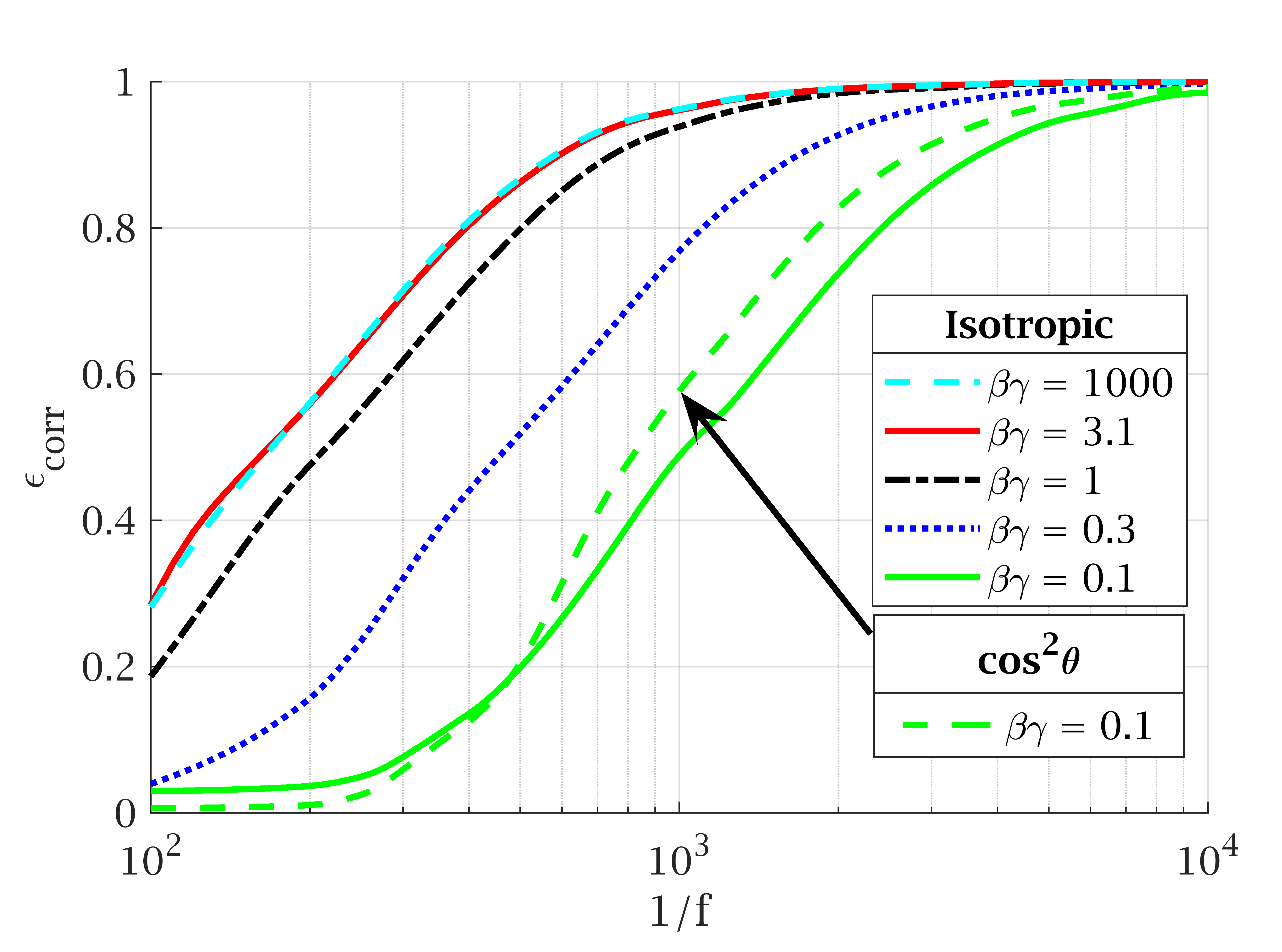}
		\caption{The LIP efficiency correction factor $\epsilon_{\tiny{corr}}(f,\beta\gamma)$ as a function of $f^{-1}$ for a variety of LIP $\beta\gamma$. This factor is multiplied with the CDMS\-lite analysis efficiency shown in Fig.~\ref{energy spectrum} to obtain the final LIP-selection efficiency. It is the smallest for $f^{-1}$\,=\,$10^{2}$ where LIPs have a higher interaction probability. It rapidly approaches unity as $f^{-1}$ increases. The statistical uncertainties are smaller than the curve thickness.}
		\label{combined_cut}
	\end{center}
\end{figure}

The single-detector-hit criterion requires the CDMS\-lite detector to be the only detector from all five stacks with a reconstructed energy deposition greater than its energy threshold~\cite{CDMSlite}. This selection criterion reduces background sources capable of depositing energy in multiple detectors. The single-detector-hit criterion is relatively efficient for LIPs with small $f$, because other detectors have energy thresholds $\sim$10 times larger ($\gtrsim1$\,keV$_{ee}$) than the CDMS\-lite detector. However, LIPs with large $f$ may deposit energy in multiple detectors and hence can be rejected by this selection criterion. To account for lost sensitivity, $\epsilon_{\tiny{\mbox{sdh}}}(f,\beta\gamma)$ is estimated as
\begin{equation}
\label{single}
\epsilon_{\tiny{\mbox{sdh}}}(f,\beta\gamma) = 1- \frac{N_{\mbox{md}}(f,\beta\gamma)}{N_{\mbox{CDMSlite}}(f,\beta\gamma)},
\end{equation}
where $N_{\mbox{CDMSlite}}(f,\beta\gamma)$ is the number of simulated LIP events depositing energy in the CDMS\-lite detector within the analyzed energy window (0.1--2\,keV), and $N_{\mbox{md}}(f,\beta\gamma)$ is the number of LIP events that also deposit energy in at least one other detector above its threshold ($\gtrsim$ 1\,keV).

Because the non-uniform electric field at high radius in the CDMS\-lite detector results in an inaccurate reconstruction of deposited energy, a fiducial-volume selection criterion was applied to remove events with energy depositions located at relatively high radius~\cite{CDMSlite}. While calculating $\epsilon_{\tiny{\mbox{fv}}}(f,\beta\gamma)$, we conservatively assume that the position reconstruction of all events with more than one interaction point in the CDMSlite detector is such that they are rejected. Hence,  
 \begin{equation}
 \label{radial}
\epsilon_{\tiny{\mbox{fv}}}(f,\beta\gamma) = 1- \frac{N_m(f,\beta\gamma)}{N_{\mbox{total}}(f,\beta\gamma)},
\end{equation}
where $N_{\mbox{total}}(f,\beta\gamma)$ is the number of simulated LIP events depositing energy in the CDMS\-lite detector, and $N_m(f,\beta\gamma)$ is the number of these interacting at more than one location in the same detector.

\begin{figure}
	\begin{center}
		\includegraphics[width=0.95\linewidth]{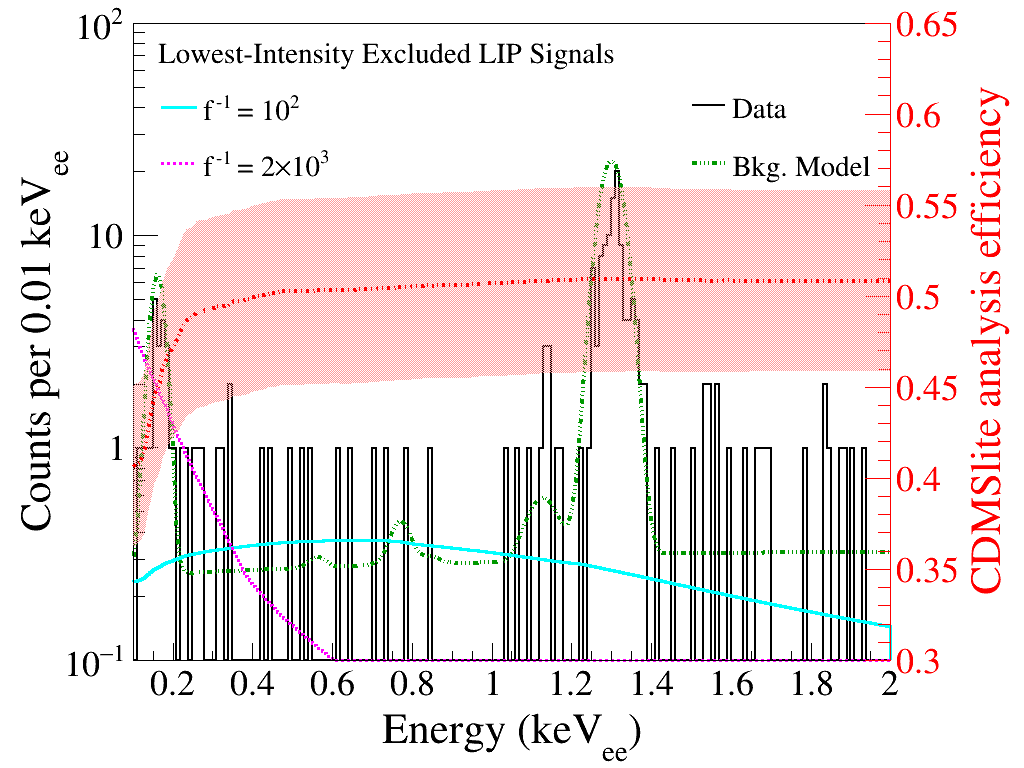}		
		\caption{The measured energy-deposition spectrum after application of all event-selection criteria (black solid histogram labeled on left axis), compared to the lowest-intensity LIP signals excluded by this analysis (see Fig.~\ref{beta_gamma_3_1}) for $f^{-1}=10^{2}$ (cyan line) and $f^{-1}=2\times10^{3}$ (dotted magenta line), and the efficiency-corrected background model (green dot-dashed curve). Also shown is the analysis efficiency (based on CDMSlite Run 2 WIMP-search~\cite{CDMSlite} and depicted by the red dashed curve labeled on right axis) with 1$\sigma$ uncertainty (red band), before the correction for additional LIP-selection inefficiency (see Fig.~\ref{combined_cut}). The energy depositions are measured in electron equivalent units (keV$_{ee}$) where it is assumed that all energy depositions in the detector are due to electron recoils~\cite{CDMSlite}.} 
		\label{energy spectrum}
	\end{center}
\end{figure}

 Figure~\ref{combined_cut} shows the combined efficiency correction factor $\epsilon_{\tiny{\mbox{corr}}}$ as a function of $f^{-1}$ for various values of $\beta\gamma$. It is smallest for $f^{-1}$\,=\,$10^{2}$ as these LIPs have a higher probability of interaction, and it rapidly approaches unity as the value of $f^{-1}$ increases. The efficiency correction was made under the approximation that the cuts were uncorrelated, which was checked to produce less than a 10\,\% inaccuracy. The correction factor is usually lower ($\lesssim$\,15\,\%) for an isotropic angular distribution than for a $\cos^2\theta$ angular distribution; an isotropic angular distribution results in a higher average LIP path length within the CDMS\-lite detector, which increases the fraction of LIPs capable of interacting more than once. The most ionizing LIPs considered ($\beta \gamma$\,$\lesssim$\,0.3 and $f^{-1}$\,$\lesssim$\,$300$) have a substantial probability of depositing above-threshold energy in the detector immediately above or below the CDMS\-lite detector, causing them to fail the single-detector-hit criterion. As a result, $\epsilon_{\tiny{\mbox{corr}}}$ is $\sim$\,3.5 times larger for the most ionizing LIPs for an isotropic distribution. The CDMS\-lite Run 2 Period 1 analysis efficiency (Fig.~\ref{energy spectrum}) is multiplied by $\epsilon_{\tiny{\mbox{corr}}}$ to obtain the final LIP-selection efficiency, $\epsilon(f,\beta\gamma,E)$.

\textit{Intensity limit calculation}: The upper limit at 90\,\% confidence level on the LIP vertical intensity, $I^{90}_v(f,\beta\gamma)$, for an isotropic incident angular distribution is given by
\begin{equation}
 \footnotesize{
I^{90}_{v}(f,\beta\gamma) = \frac{N^{90}(f,\beta\gamma)}{\tau \displaystyle\int_{0.1\,keV}^{2\,keV}\epsilon(f,\beta\gamma,E)\int  \frac{\mathrm{d}P}{\mathrm{d}E}(f,\beta\gamma,\theta)A(\theta)\,\mathrm{d}\Omega\mathrm{d}E}},
\label{intensity}
\end{equation}
where $N^{90}(f,\beta\gamma)$ is the 90\,\% confidence upper limit on the expected number of observed LIPs, $\tau$ is the live time of the detector, $\frac{\mathrm{d}P}{\mathrm{d}E}(f, \beta\gamma,\theta)$ is the LIP energy-deposition distribution at LIP incident angle $\theta$ and $\epsilon(f,\beta\gamma,E)$ is the LIP-selection efficiency. The effective cross-sectional area of the detector surface at $\theta$ is $A(\theta) = \pi r^2\cos\theta + 2rh\sin\theta,$ where $r$ and $h$ are the detector radius and height, respectively. To compute $I^{90}_{v}(f,\beta\gamma)$ for a $\cos^2\theta$ angular distribution, $\frac{\mathrm{d}P}{\mathrm{d}E}(f, \beta\gamma,\theta)$ is weighted by a $\cos^2\theta$ factor. 
    
We calculate $N^{90}(f,\beta\gamma)$ using the Optimum Interval (OI) method~\cite{yellin2002finding} under the conservative assumption that all observed events in the energy-deposition distribution could be due to LIP interactions. This method does not provide any discovery potential. The values of $N^{90}(f)$ obtained are between 41 and 79.

\textit{Expected sensitivity and uncertainty}: The LIPs projected sensitivity is determined by computing the mean expected upper limit from background alone, based on 200 different energy-deposition spectra simulated using the CDMSlite Run 2 background model~\cite{titrium}. Each simulated distribution contains a random number of events that is statistically consistent with that predicted for the Period 1 live time. For each sensitivity calculation, the analysis efficiency and the energy thresholds are varied within their uncertainties. The resulting 1$\sigma$ uncertainty in the sensitivity is $\sim$32\,\%.

It is difficult to estimate the systematic uncertainty due to possible deviation of the true $\langle\mathrm{d}P/\mathrm{d}E\rangle$ from that given by \textsc{Geant4}. We estimate this uncertainty by comparing the sensitivity obtained herein with that resulting from $\langle\mathrm{d}P/\mathrm{d}E\rangle$ obtained using the CDMS II convolution method~\cite{prasad_2013,cdmsii-lips}.  Our \textsc{Geant4}-based sensitivity is $\sim$24\,\% less restrictive, which we take as an estimate of the systematic uncertainty on $\langle\mathrm{d}P/\mathrm{d}E\rangle$ for the entire range of $\beta\gamma$ values considered.  We estimate the total uncertainty ($\sim$40\,\%) on the expected sensitivity by combining this $\langle\mathrm{d}P/\mathrm{d}E\rangle$ systematic uncertainty in quadrature with the estimate for the other sources of uncertainty. The uncertainty on the final LIP vertical intensity limit is $\sim$37\,\%; it includes the analysis-efficiency and energy-threshold uncertainties, and the $\langle\mathrm{d}P/\mathrm{d}E\rangle$ systematic uncertainty.

\textit{Unblinding and results}: We examined the LIP-search data for the first time only after finalizing the event-selection criteria and their efficiencies, the systematic uncertainties, and the procedure for calculating the LIP vertical intensity. The measured spectrum contains 180 events after application of all selection criteria and is shown in Fig.~\ref{energy spectrum}. The most prominent features in the spectrum are the L- and M-shell peaks from decays of intrinsic Ge radioisotopes, as described in Ref.~\cite{CDMSlite}.
A general agreement was observed between the data spectrum and the simulated background spectrum.

\begin{figure}
	\begin{center}
		\includegraphics[width=240pt]{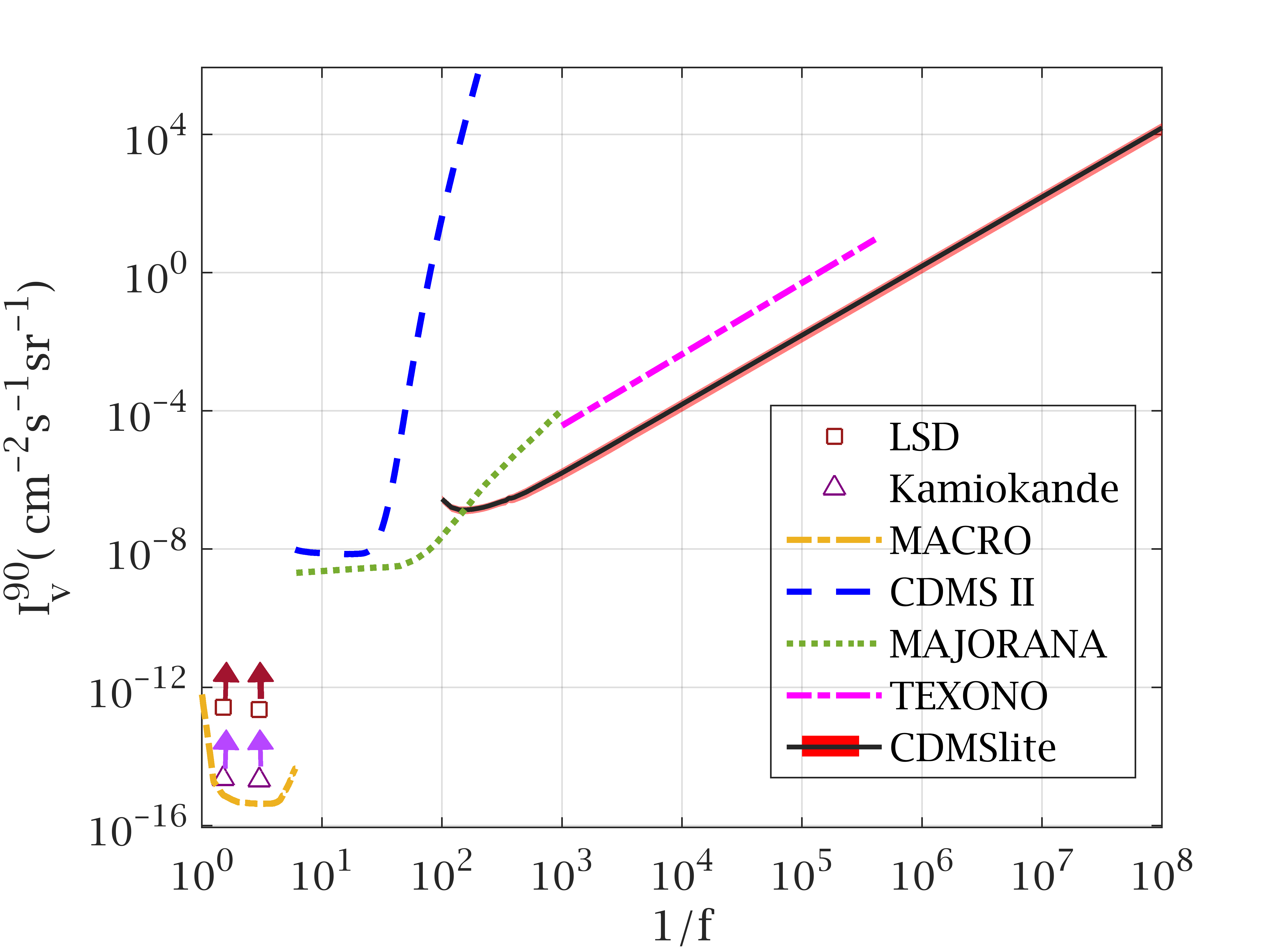}
	        \caption{The CDMSlite 90\,\% upper confidence limit on the LIP vertical intensity (solid black) under the assumptions of an isotropic distribution for minimum-ionizing LIPs. The red band shows the 1$\sigma$ level uncertainty  in the limit. The result is compared to those from all prior searches for cosmogenic LIPs, including  LSD~\cite{LSD} (brown $\square$), Kamiokande~\cite{kamiokandee} (purple $\Delta$), MACRO~\cite{MACRO} (yellow dot-dashed), CDMS-II~\cite{cdmsii-lips} (blue dashed), \textsc{Majorana}~\cite{Majorana} (green dotted), and TEXONO~\cite{texono} (magenta dot-dashed). The expected sensitivity generally lies on top of the final limit to within the resolution of the plot except for a small range of larger fractional charges ($160$\,$<$\,$f^{-1}$\,$<$\,$500$), where the limits are slightly less restrictive.} 
		\label{beta_gamma_3_1}
	\end{center}
\end{figure}

Due to the $f^2$ suppression of the interaction rate in the rock overburden, the expected intensity of energetic LIPs ($\beta\gamma$\,$\geq$\,3.1) at the experiment is minimally reduced relative to that at the surface for the range of LIP charges considered. LIP $\beta\gamma$ is reduced by $\lesssim$10\,\% for LIPs with $f^{-1} > 10^4$, and for LIPs with $\beta\gamma\geq1$ and $f^{-1} > 10^3$. LIPs with mass $\lesssim$1\,GeV/$c^2$, lower values of $\beta\gamma$, and lower values of $f^{-1}$ may be attenuated or have their value of $\beta\gamma$ reduced by the overburden. The vertical-intensity limit $I^{90}_v(f,\beta\gamma)$ is shown in Fig.~\ref{beta_gamma_3_1} for a minimum-ionizing LIP ($\beta\gamma$\,=\,3.1) with an isotropic incident distribution and is compared to limits from prior direct searches for cosmogenic LIPs.
This result sets the strongest constraint on LIPs with $f^{-1}$\,$>$\,160,
including a minimum vertical-intensity  limit of $1.36\times10^{-7}$\,cm$^{-2}$s$^{-1}$sr$^{-1}$ at $f^{-1}$\,$=$\,160. The final limit agrees with the expected sensitivity to within about 2\,$\sigma$ for $160$\,$<$\,$f^{-1}$\,$<500$ and within 1\,$\sigma$ elsewhere.

\begin{figure}
	\begin{center}
		\includegraphics[width=240pt]{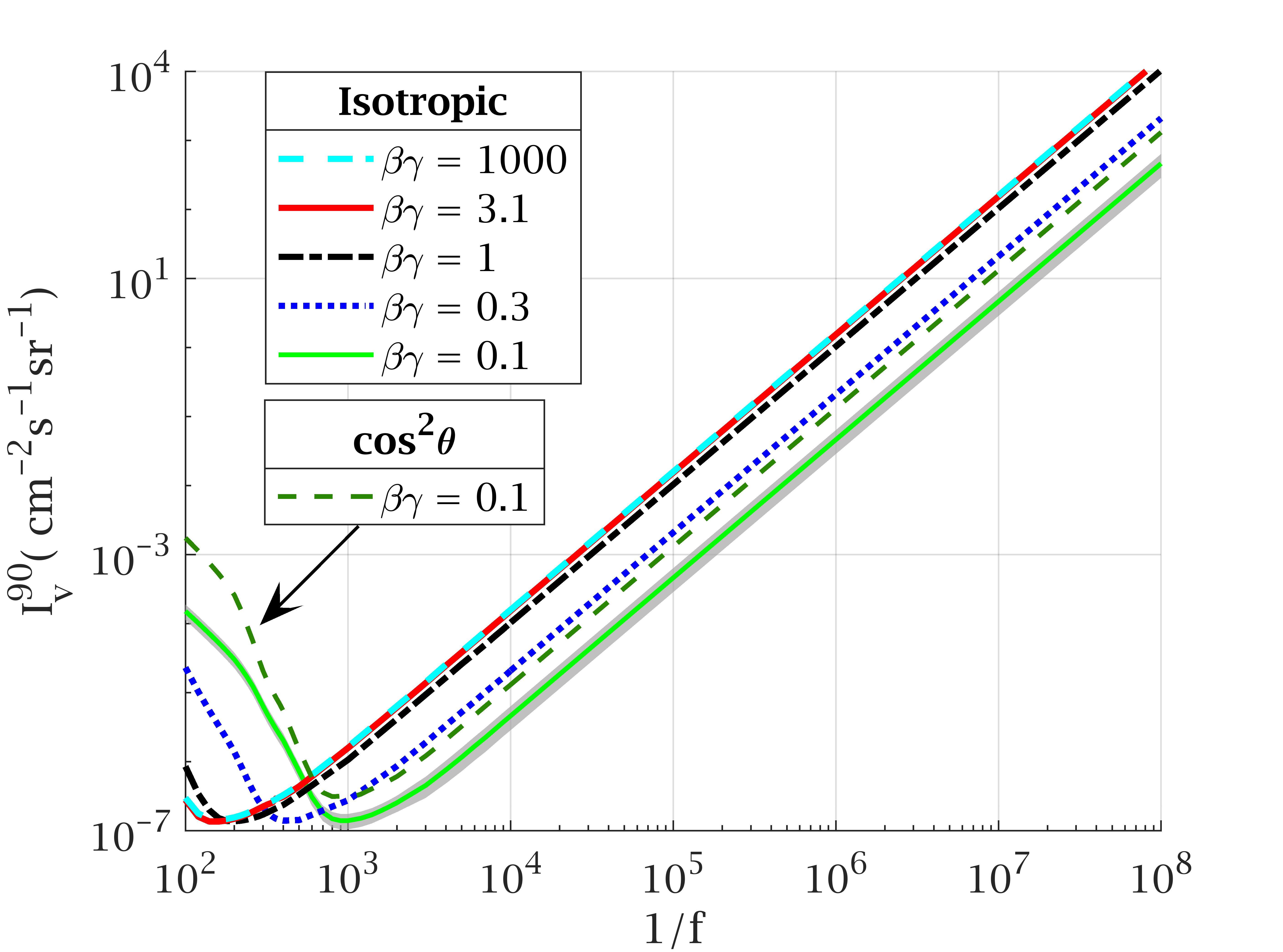}
		\caption{The 90\,\% confidence limit on LIP vertical intensity as a function of LIP electric charge for various values of LIP $\beta\gamma$. The limit curves for $\beta\gamma \geq 10^3$ coincide with each other and are represented by a single curve. For clarity, the uncertainty band (light gray) is only shown for $\beta\gamma$\,=\,0.1 but is indicative of the size of the uncertainty of all the limit curves. The $\beta\gamma$\,=\,3.1 curve is the same as that in Fig.~\ref{beta_gamma_3_1}. } 
		\label{sens_all_isotropic}
	\end{center}
\end{figure}

Figure~\ref{sens_all_isotropic} shows the limits for a variety of $\beta\gamma$ values. The results are valid for the entire mass range considered: 5\,MeV/$c^2$ to 100\,TeV/$c^2$. The intensity limit computed for a cos$^2\theta$ angular distribution is nearly three times weaker than that for an isotropic angular distribution for most values of $f$. 

\textit{Summary}: Utilizing a SuperCDMS detector operated in CDMS\-lite mode, this work presents the first direct-detection limits on the vertical intensity of cosmogenic LIPs with charge less than $e/(3\times10^5$) for values of incident $\beta\gamma$ ranging from 0.1 to 10$^6$.
Although the OI limit-setting method used does not have discovery potential, the result reported herein represents a significant step towards searching for dark matter with fractional charge~\cite{annualMod, annualMod2} by setting the first limit on non-relativistic LIPs with $\beta\gamma$ values as small as 0.1. Future searches extending to yet lower values of $\beta\gamma$ may probe galactically bound LIPs.

\textit{Acknowledgments}: The SuperCDMS collaboration gratefully acknowledges technical assistance from the staff of the Soudan Underground Laboratory and the Minnesota Department of Natural Resources. The CDMS\-lite and iZIP detectors were fabricated in the Stanford Nanofabrication Facility, which is a member of the National Nanofabrication Infrastructure Network, sponsored and supported by the NSF. Funding and support were received from the National Science Foundation, the U.S.\ Department of Energy (DOE), NSF OISE 1743790, Fermilab URA Visiting Scholar Grant No.\ 15-S-33, NSERC Canada, the Canada First Excellence Research  Fund, the Arthur B. McDonald Institute (Canada),  the Department of Atomic Energy Government of India (DAE), the Department of Science and Technology (DST, India) and the DFG (Germany) - Project No.\ 420484612 and under Germany's Excellence Strategy - EXC 2121 ``Quantum Universe" - 390833306. Femilab is operated by Fermi Research Alliance, LLC,  SLAC is operated by Stanford University, and PNNL is operated by the Battelle Memorial Institute for the U.S.\ Department of Energy under contracts DE-AC02-37407CH11359, DE-AC02-76SF00515, and DE-AC05-76RL01830, respectively.

\bibliographystyle{apsrev4-1}
\bibliography{bibliography.bib}

\end{document}